# Cited Half-Life of the Journal Literature

Philip M. Davis, Independent Researcher and Publishing Consultant
Phil Davis Consulting
pmd8@cornell.edu
orcid.org/0000-0002-1893-5360
(corresponding author)

Angela Cochran
Director, Journals
American Society of Civil Engineers
orcid.org/0000-0002-6255-4389

April 29, 2015

*Correction note:* Table 1 includes is replaced to fix a sorting error.

## Abstract

Analyzing 13,455 journals listed in the *Journal Citation Report* (Thomson Reuters) from 1997 through 2013, we report that the mean cited half-life of the scholarly literature is 6.5 years and growing at a rate of 0.13 years per annum. Focusing on a subset of journals (N=4,937) for which we have a continuous series of half-life observations, 209 of 229 (91%) subject categories experienced increasing cited half-lives. Contrary to the overall trend, engineering and chemistry journals experienced declining cited half-lives. Last, as journals attracted more citations, a larger proportion of them were directed toward older papers. The trend to cite older papers is not fully explained by technology (digital publishing, search and retrieval, etc.), but may be the result of a structural shift to fund incremental and applied research over fundamental science.

## Introduction

Are scholars citing older materials? Analyzing the age of citations in a corpus of literature indexed by Google Scholar, from 1990 through 2013, Verstak and others[1] computed the proportion of citations to "older articles," defined as being 10 or more years old at the time of citation. They reported increasing half-lives for 7 out of 9 broad subject categories, the two exceptions being Chemical & Material Sciences and Engineering.

This study attempts to validate Verstak using a different methodological approach. Rather than relying on the individual *citation* as the unit of observation, we base our analysis on the cited half-life of *journals*, as reported in the *Journal Citation Report* (Thomson Reuters). This approach has the obvious advantage of scale, allowing us to approach the problem using thousands of journals rather than tens of millions of citations.

## Dataset

The dataset includes 13,455 unique journal names as reported in Thomson Reuter's *Journal Citation Report* (*JCR*) from 1997 through 2013. The primary variable of interest in this analysis is the *cited half-life*, which is the median age of articles cited in a given journal for a given year. By definition, half of the articles in a journal will be older than the cited half-life; the other half will be younger.

Of the 13,455 unique journal names listed in our dataset, 4,937 (37%) received a continuous series of half-life calculations over the 17-year dataset. We will refer to this subset as the "old set." The remaining dataset consists of journals that were added to the *JCR* after 1997, changed names, or were delisted at some point. We will refer to this group as the "new set," and the entire dataset as the "full set." We make these distinctions to attempt to isolate the behavioral effects of citing authors, from technical changes to the *JCR* dataset.

The full dataset includes a total of 197,997 journal half-life observations, classified into 232 subject categories, with some journals receiving multiple classifications. The dataset grew from 6,438 journals in 1997 to 11,022 in 2013, with many new journals added between 2008 and 2010 (Figure 1).

While subject categories were relatively stable, some new subject categories were added throughout the timeframe. For example, NANOSCIENCE & NANOTECHNOLOGY began as a distinct journal category in 2006. Other subjects (e.g. BIOLOGY, MISCELLANEOUS) were discontinued as distinct subject categories.

The size of each journal was measured by the number of "citable items" published in a given year. Citable items include papers classified by Thomson Reuters as Research Articles, Review Articles and Proceedings Papers. Editorials, Letters to the Editor, Corrections, and other article types were excluded from the citable item count.

Lastly, nearly 17.5% of cited half-life observations were listed in the dataset as >10, meaning a half-life greater than 10 years. These observations were recoded as 11 so that they could be used in the analysis. To evaluate the effect of this recoding on the results of the analyses, we report the results when >10 observations were replaced with missing values.

**Statistical Analysis**

This study is based on a linear, multi-level model with repeated measures. The variable of interest (our response variable) was the CITED HALF-LIFE for each journal. The covariates in this study were the YEAR of observation (1997… 2013), the JOURNAL, and its subject CATEGORY (or categories). Each journal was measured repeatedly (each year) over the longitudinal dataset and was nested within its subject CATEGORY (or categories, if there were multiple). In order to measure whether the cited half-life of subject categories changed over time, we created an interaction variable between CATEGORY and TIME.

Lastly, as journal size varies considerably across our dataset (for example, *PLOS ONE* published more than 30,000 Research Articles in 2013), we weighted our regression model with the number of CITABLE ITEMS published in each journal.

All analyses were performed using JMP v11 (SAS).

**Results**

For the full set of journals, the mean weighted cited half-life was 6.5 years, growing at a rate of 0.13 years per annum ($p<.0001$). Mean cited half-life for the old set of journals was 7.1 years, growing at a rate of 0.13 years per annum ($p<.0001$), and mean cited half-life for the new set of journals was 5.1 years, growing at a rate of 0.19 years per annum ($p<.0001$). When >10 year observations were replaced with missing values, growth rates were similar (0.13 for the full set, 0.12 for the old set, and 0.20 for the new set).

Focusing on the old set of journals (N=4,937) for which we have a continuous series of cited half-life observations, 91% (209 of 229) journal categories experienced increasing cited half-lives. For example, the cited half-life for journals classified under DEVELOPMENTAL BIOLOGY grew at 0.25 years per annum (95% C.I. 0.20 to 0.29, $p<.0001$), GENETICS & HEREDITY grew at 0.20 years per annum (95% C.I. 0.17 to 0.23, $p<.0001$), and CELL BIOLOGY grew at 0.17 years per annum (95% C.I. 0.15 to 0.20, $p<.0001$), Table 1.

Conversely, the cited half-life of just 20 (9%) journal categories decreased over our observation period. With few exceptions, these subjects covered the general fields of chemistry and engineering. For example, the cited half-life for journals classified under ENERGY & FUELS declined by 0.11 years per annum (95% C.I. -0.15 to -0.08, $p<.0001$),

CHEMISTRY, MULTIDISCIPLINARY declined by 0.07 years per annum (95% C.I. -0.10 to -0.05, p<.0001), ENGINEERING, MULTIDISCIPLINARY declined by 0.05 years per annum (95% C.I. -0.10 to -0.1, p<.0001), and ENGINEERING, CHEMICAL declined by 0.04 years per annum (95% C.I. -0.07 to -0.02, p<.0001).

Lastly, cited half-life increased with total citations. For every logTOTAL CITATIONS (about 2.7 citations), the cited journal half-life increased by about 0.4%. In other words, as a journal attracted more citations, a larger proportion of them were directed towards older papers. From a distribution perspective, more citations shifts the age of cited journal references to the right. This dynamic can be visualized in Figure 3 as journals move into the top-right quadrant in the graph.

A dynamic bubble plot animating the change in cited half-life for categories and journals can be viewed at http://bit.ly/1yTku2n (.swf file). Categories may be split to reveal the path of individual journals. (Mac users may need to hold the Control button and select their browser application to open the file, or change their browser's security preferences to open .swf files).

## Discussion

Similar to Verstak[1] and Larivière,[2] we report a general increase in the cited half-life of the scholarly literature, and like Verstak, we report that some fields (viz. chemistry and engineering) are moving toward shorter half-lives. In addition, we find that cited half-life is related to total citations, meaning, that as a journal attracts more citations, a larger proportion of them are citing older literature.

Explaining why cited half-life of citations may be increasing, Vertstak[1] lists several changes in the way scholarly information is produced, disseminated, and discovered, including:

1. A industry-wide transformation from print to online distribution,
2. Mass digitization of backfiles and journal archives,
3. Full text indexing, and
4. Better search engines based on relevance ranking.

The increase in cited half-life has been documented back to the 1960s,[2] decades before digital publishing and complementary tools were developed. While these tools may help authors discover and retrieve relevant literature, they do not help to explain a decades-long trend that began during printed publication. Indeed, technologies like the photocopier, fax, and email may provide a more adequate explanation for growth from the 1970s.

In his seminal paper, Networks of Scientific Papers,[3] Price described how the current literature is "knitted" to just a small percentage of older papers. He calls this selective reference to the older literature an "active research front." At the same time, Price provides empirical evidence for *obsolescence,* in which older materials cease to be cited. Obsolescence may be the result of the declining relevance of a scientific contribution or because that contribution has become fully incorporated into the literature and does not require further reference. This idea was first described by Robert Merton in 1949[4] but promoted as "obliteration by incorporation" by Eugene Garfield.[5]

A growing half-life of the journal literature reflects many cultural, technological, and normative behaviors of citing authors all acting in concert. Neither Verstak[1] nor Larivière[6] provide adequate explanations for why authors are citing older material, or why some fields, like chemistry or engineering are moving in the opposite direction.

In addition to new tools that aid in the production, dissemination, discovery, and retrieval of scientific results, a growing cited half-life might reflect major structural shifts in the way science is funded and the way scientists are rewarded. A move to fund incremental and applied research may result in fewer fundamental and theoretical studies being published. Giving credit to these founders may require that authors cite an increasingly aging literature.

**Acknowledgements**
We would like to thank Stephen Mobley, Patricia Brennan and the RD&S Custom Data Team at Thomson Reuters Web of Science for proving us with the data.

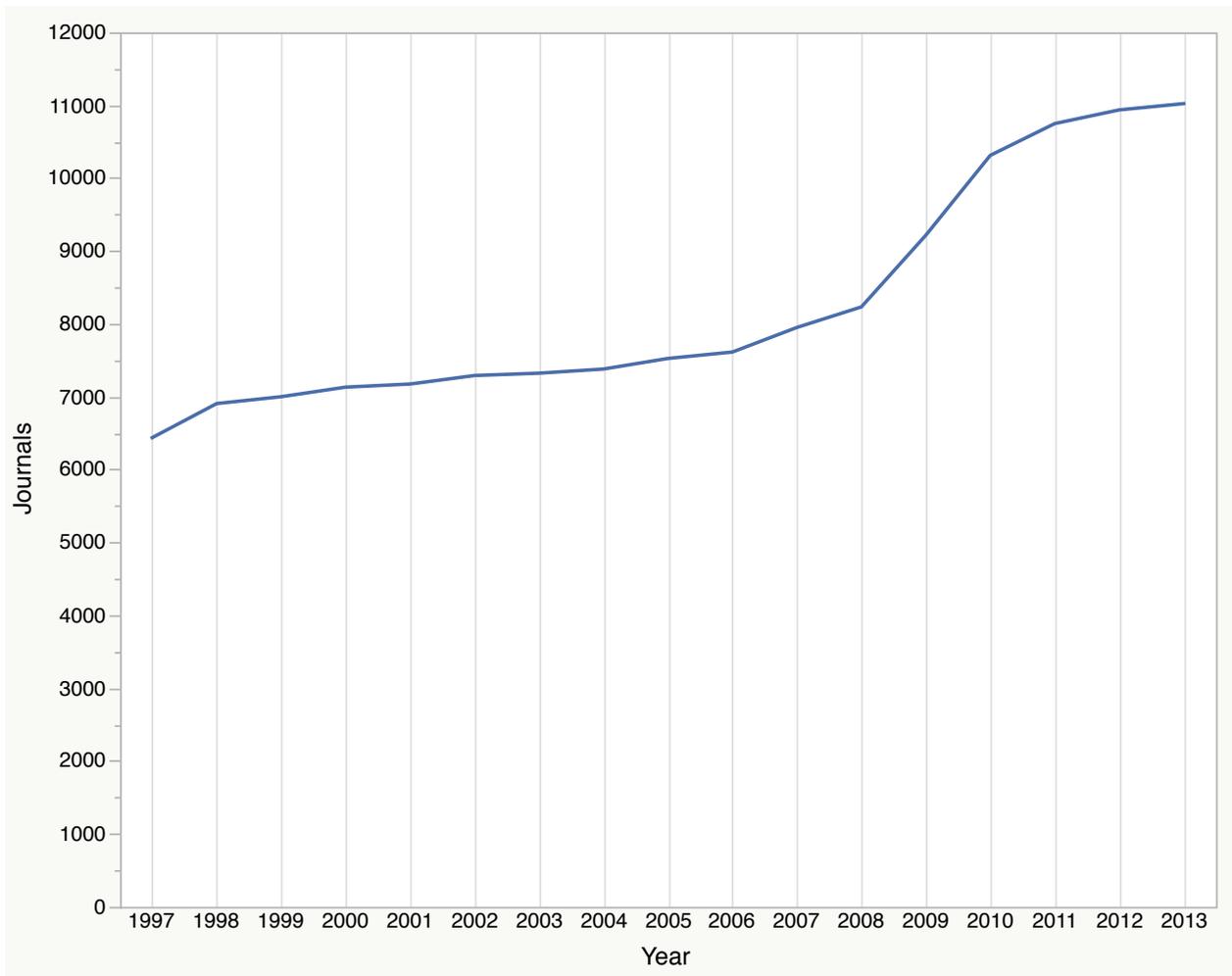

**Figure 1.** Number of journals indexed in the *Journal Citation Report,* 1997—2013.

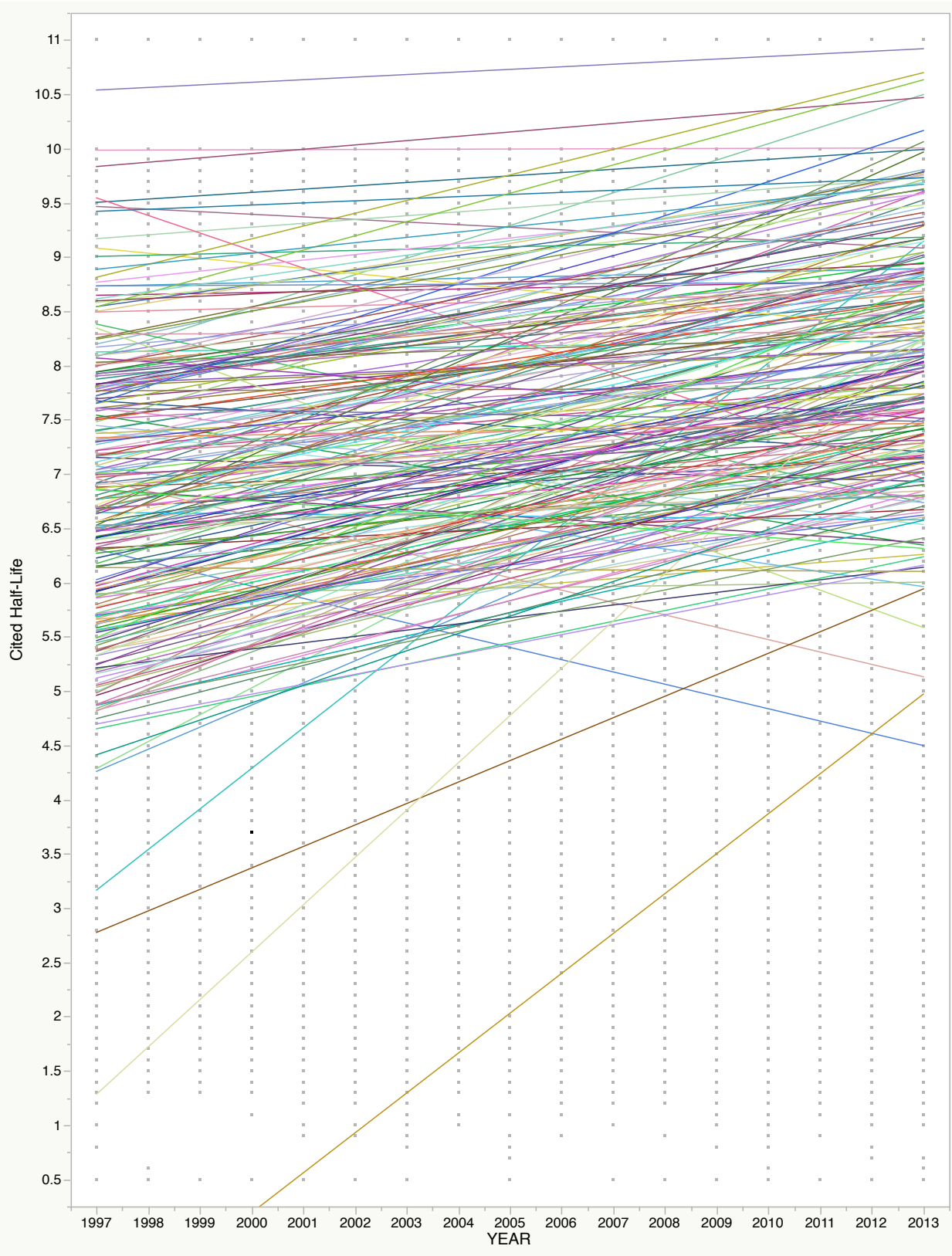

**Figure 2.** Cited half-life growth estimates for 232 journal subject categories, 1997—2013.

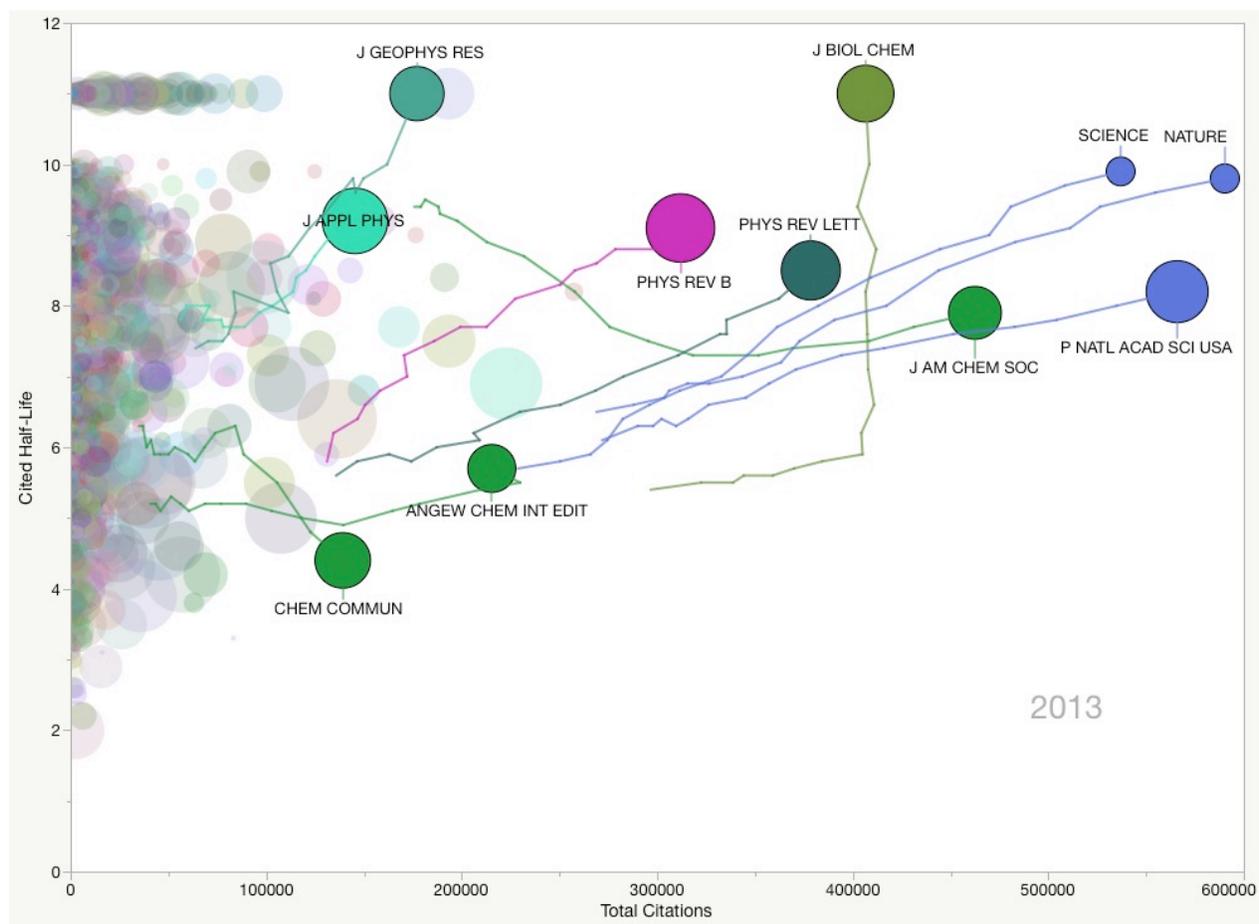

**Figure 3.** Bubble plot illustrating the positive relationship between total citations and cited half-life, 1997—2013. Journals are sized by citable items published in a given year and colored by category. The paths of highly-cited journals are highlighted. A dynamic version of this graph can downloaded from http://bit.ly/1biSt9j (.swf file type).

**Table 1.** Mean cited half-life and growth estimates by subject category.

| CATEGORY | Journals | Mean Cited Half-Life | Growth per Annum |
|---|---|---|---|
| ACOUSTICS | 303 | 7.65 | 0.06 |
| AGRICULTURAL ECONOMICS & POLICY | 125 | 7.39 | 0.07 |
| AGRICULTURAL ENGINEERING | 67 | 6.38 | -0.11 |
| AGRICULTURE, DAIRY & ANIMAL SCIENCE | 457 | 7.70 | 0.04 |
| AGRICULTURE, MULTIDISCIPLINARY | 224 | 8.33 | 0.09 |
| AGRONOMY | 704 | 8.12 | 0.07 |
| ALLERGY | 238 | 5.83 | 0.04 |
| ANATOMY & MORPHOLOGY | 210 | 8.38 | 0.15 |
| ANDROLOGY | 51 | 7.50 | 0.05 |
| ANESTHESIOLOGY | 254 | 6.54 | 0.12 |
| ANTHROPOLOGY | 657 | 9.51 | 0.01 |
| AREA STUDIES | 446 | 8.60 | 0.18 |
| ASTRONOMY & ASTROPHYSICS | 553 | 7.56 | 0.02 |
| AUDIOLOGY & SPEECH-LANGUAGE PATHOLOGY | 33 | 9.70 | 0.15 |
| AUTOMATION & CONTROL SYSTEMS | 490 | 7.53 | -0.03 |
| BEHAVIORAL SCIENCES | 597 | 8.33 | 0.04 |
| BIOCHEMICAL RESEARCH METHODS | 461 | 7.23 | 0.09 |
| BIOCHEMISTRY & MOLECULAR BIOLOGY | 2928 | 6.63 | 0.16 |
| BIODIVERSITY CONSERVATION | 167 | 9.24 | 0.01 |
| BIOLOGY | 693 | 8.12 | 0.17 |
| BIOLOGY, MISCELLANEOUS | 257 | 7.78 | 0.04 |
| BIOPHYSICS | 576 | 7.39 | 0.15 |
| BIOTECHNOLOGY & APPLIED MICROBIOLOGY | 1245 | 6.65 | 0.13 |
| BUSINESS | 748 | 8.91 | 0.16 |
| BUSINESS, FINANCE | 416 | 8.96 | 0.13 |
| CARDIAC & CARDIOVASCULAR SYSTEMS | 792 | 6.12 | 0.06 |
| CELL & TISSUE ENGINEERING | 9 | 4.23 | 0.37 |
| CELL BIOLOGY | 1694 | 6.48 | 0.17 |
| CHEMISTRY, ANALYTICAL | 724 | 6.71 | 0.06 |
| CHEMISTRY, APPLIED | 504 | 7.49 | -0.01 |
| CHEMISTRY, INORGANIC & NUCLEAR | 466 | 7.88 | 0.04 |
| CHEMISTRY, MEDICINAL | 375 | 6.41 | 0.08 |
| CHEMISTRY, MULTIDISCIPLINARY | 1106 | 7.24 | -0.07 |
| CHEMISTRY, ORGANIC | 593 | 7.34 | 0.09 |
| CHEMISTRY, PHYSICAL | 1177 | 7.32 | 0.11 |
| CLINICAL NEUROLOGY | 1497 | 6.89 | 0.08 |
| COMMUNICATION | 474 | 8.80 | 0.10 |
| COMPUTER SCIENCE, ARTIFICIAL INTELLIGENCE | 795 | 6.91 | 0.10 |
| COMPUTER SCIENCE, CYBERNETICS | 251 | 7.13 | 0.11 |
| COMPUTER SCIENCE, HARDWARE & ARCHITECTURE | 481 | 7.85 | 0.05 |
| COMPUTER SCIENCE, INFORMATION SYSTEMS | 623 | 7.22 | 0.07 |
| COMPUTER SCIENCE, INTERDISCIPLINARY APPLICATIONS | 731 | 6.97 | 0.02 |
| COMPUTER SCIENCE, SOFTWARE ENGINEERING | 699 | 7.80 | 0.07 |
| COMPUTER SCIENCE, THEORY & METHODS | 775 | 8.12 | 0.07 |

Table 1. (con't)

| CATEGORY | Journals | Mean Cited Half-Life | Growth per Annum |
|---|---|---|---|
| CONSTRUCTION & BUILDING TECHNOLOGY | 237 | 7.45 | 0.03 |
| CRIMINOLOGY & PENOLOGY | 255 | 8.00 | 0.03 |
| CRITICAL CARE MEDICINE | 183 | 6.17 | 0.11 |
| CRYSTALLOGRAPHY | 280 | 8.19 | 0.13 |
| CULTURAL STUDIES | 13 | 8.98 | 0.37 |
| DEMOGRAPHY | 204 | 8.84 | 0.08 |
| DENTISTRY, ORAL SURGERY & MEDICINE | 624 | 8.58 | 0.01 |
| DERMATOLOGY | 497 | 7.34 | 0.10 |
| DEVELOPMENTAL BIOLOGY | 345 | 6.58 | 0.25 |
| ECOLOGY | 1205 | 8.24 | 0.08 |
| ECONOMICS | 2331 | 8.69 | 0.09 |
| EDUCATION & EDUCATIONAL RESEARCH | 1353 | 8.42 | 0.08 |
| EDUCATION, SCIENTIFIC DISCIPLINES | 173 | 7.67 | 0.02 |
| EDUCATION, SPECIAL | 270 | 8.70 | -0.10 |
| ELECTROCHEMISTRY | 164 | 6.46 | -0.07 |
| EMERGENCY MEDICINE | 158 | 6.20 | 0.12 |
| ENDOCRINOLOGY & METABOLISM | 1133 | 6.21 | 0.10 |
| ENERGY & FUELS | 635 | 7.18 | -0.11 |
| ENGINEERING, AEROSPACE | 315 | 8.27 | 0.09 |
| ENGINEERING, BIOMEDICAL | 470 | 6.88 | 0.08 |
| ENGINEERING, CHEMICAL | 1266 | 7.57 | -0.04 |
| ENGINEERING, CIVIL | 621 | 7.53 | -0.03 |
| ENGINEERING, ELECTRICAL & ELECTRONIC | 2228 | 6.83 | 0.11 |
| ENGINEERING, ENVIRONMENTAL | 343 | 6.25 | 0.01 |
| ENGINEERING, GEOLOGICAL | 215 | 8.47 | 0.06 |
| ENGINEERING, INDUSTRIAL | 366 | 7.81 | 0.13 |
| ENGINEERING, MANUFACTURING | 288 | 6.60 | 0.13 |
| ENGINEERING, MARINE | 79 | 8.79 | 0.00 |
| ENGINEERING, MECHANICAL | 1155 | 7.36 | 0.05 |
| ENGINEERING, MULTIDISCIPLINARY | 608 | 7.71 | -0.05 |
| ENGINEERING, OCEAN | 112 | 7.79 | 0.13 |
| ENGINEERING, PETROLEUM | 151 | 9.20 | -0.02 |
| ENTOMOLOGY | 897 | 9.22 | 0.08 |
| ENVIRONMENTAL SCIENCES | 1520 | 6.96 | 0.05 |
| ENVIRONMENTAL STUDIES | 664 | 7.45 | 0.00 |
| ERGONOMICS | 199 | 7.83 | 0.06 |
| ETHICS | 295 | 8.30 | 0.08 |
| ETHNIC STUDIES | 68 | 8.15 | 0.13 |
| EVOLUTIONARY BIOLOGY | 331 | 8.20 | 0.14 |
| FAMILY STUDIES | 364 | 8.52 | 0.04 |
| FISHERIES | 439 | 8.15 | 0.12 |
| FOOD SCIENCE & TECHNOLOGY | 1020 | 8.03 | 0.03 |
| FORESTRY | 376 | 8.68 | 0.11 |

Table 1. (con't)

| CATEGORY | Journals | Mean Cited Half-Life | Growth per Annum |
|---|---|---|---|
| GASTROENTEROLOGY & HEPATOLOGY | 540 | 5.78 | 0.12 |
| GENETICS & HEREDITY | 1156 | 7.24 | 0.20 |
| GEOCHEMISTRY & GEOPHYSICS | 596 | 8.94 | 0.07 |
| GEOGRAPHY | 589 | 7.62 | 0.06 |
| GEOGRAPHY, PHYSICAL | 343 | 8.02 | 0.04 |
| GEOLOGY | 343 | 8.66 | 0.21 |
| GEOSCIENCES, MULTIDISCIPLINARY | 1388 | 8.21 | 0.11 |
| GERIATRICS & GERONTOLOGY | 395 | 6.38 | 0.06 |
| GERONTOLOGY | 305 | 7.26 | 0.17 |
| HEALTH CARE SCIENCES & SERVICES | 532 | 7.08 | 0.09 |
| HEALTH POLICY & SERVICES | 515 | 6.67 | 0.10 |
| HEMATOLOGY | 768 | 5.87 | 0.11 |
| HISTORY | 222 | 10.73 | 0.02 |
| HISTORY & PHILOSOPHY OF SCIENCE | 458 | 10.08 | 0.00 |
| HISTORY OF SOCIAL SCIENCES | 258 | 9.31 | 0.10 |
| HORTICULTURE | 234 | 8.80 | 0.06 |
| HOSPITALITY, LEISURE, SPORT & TOURISM | 71 | 10.16 | 0.10 |
| IMAGING SCIENCE & PHOTOGRAPHIC TECHNOLOGY | 123 | 7.18 | 0.16 |
| IMMUNOLOGY | 1401 | 6.22 | 0.12 |
| INDUSTRIAL RELATIONS & LABOR | 213 | 8.22 | 0.16 |
| INFECTIOUS DISEASES | 479 | 5.67 | 0.12 |
| INFORMATION SCIENCE & LIBRARY SCIENCE | 584 | 6.94 | 0.20 |
| INSTRUMENTS & INSTRUMENTATION | 509 | 7.15 | 0.13 |
| INTEGRATIVE & COMPLEMENTARY MEDICINE | 42 | 7.92 | 0.02 |
| INTERNATIONAL RELATIONS | 598 | 6.89 | 0.13 |
| LANGUAGE & LINGUISTICS | 104 | 8.95 | -0.17 |
| LAW | 1412 | 8.09 | 0.14 |
| LIMNOLOGY | 136 | 7.86 | 0.16 |
| LINGUISTICS | 469 | 9.54 | 0.11 |
| LOGIC | 18 | 8.94 | 0.17 |
| MANAGEMENT | 917 | 8.67 | 0.14 |
| MARINE & FRESHWATER BIOLOGY | 908 | 8.56 | 0.08 |
| MATERIALS SCIENCE, BIOMATERIALS | 119 | 6.32 | 0.06 |
| MATERIALS SCIENCE, CERAMICS | 151 | 7.80 | 0.05 |
| MATERIALS SCIENCE, CHARACTERIZATION & TESTING | 240 | 7.80 | 0.03 |
| MATERIALS SCIENCE, COATINGS & FILMS | 145 | 6.68 | 0.12 |
| MATERIALS SCIENCE, COMPOSITES | 170 | 7.15 | 0.02 |
| MATERIALS SCIENCE, MULTIDISCIPLINARY | 1376 | 7.05 | 0.11 |
| MATERIALS SCIENCE, PAPER & WOOD | 167 | 10.22 | 0.04 |
| MATERIALS SCIENCE, TEXTILES | 97 | 9.36 | -0.05 |
| MATHEMATICAL & COMPUTATIONAL BIOLOGY | 106 | 9.18 | 0.04 |
| MATHEMATICS | 2056 | 9.91 | 0.03 |
| MATHEMATICS, APPLIED | 1664 | 8.51 | 0.02 |
| MATHEMATICS, INTERDISCIPLINARY APPLICATIONS | 518 | 9.23 | -0.16 |

Table 1. (con't)

| CATEGORY | Journals | Mean Cited Half-Life | Growth per Annum |
|---|---|---|---|
| MECHANICS | 1201 | 8.13 | 0.02 |
| MEDICAL ETHICS | 56 | 6.12 | 0.10 |
| MEDICAL INFORMATICS | 159 | 6.57 | 0.16 |
| MEDICAL LABORATORY TECHNOLOGY | 241 | 7.42 | -0.03 |
| MEDICINE, GENERAL & INTERNAL | 1190 | 7.18 | 0.04 |
| MEDICINE, LEGAL | 182 | 6.65 | 0.12 |
| MEDICINE, RESEARCH & EXPERIMENTAL | 697 | 6.74 | 0.05 |
| METALLURGY & METALLURGICAL ENGINEERING | 528 | 8.37 | 0.11 |
| METEOROLOGY & ATMOSPHERIC SCIENCES | 549 | 7.94 | 0.07 |
| MICROBIOLOGY | 928 | 6.91 | 0.17 |
| MICROSCOPY | 136 | 7.05 | 0.23 |
| MINERALOGY | 256 | 9.31 | 0.06 |
| MINING & MINERAL PROCESSING | 170 | 7.98 | 0.13 |
| MULTIDISCIPLINARY SCIENCES | 479 | 8.24 | 0.14 |
| MYCOLOGY | 197 | 7.88 | 0.10 |
| NANOSCIENCE & NANOTECHNOLOGY | 87 | 5.61 | 0.20 |
| NEUROIMAGING | 114 | 6.12 | 0.10 |
| NEUROSCIENCES | 2242 | 6.81 | 0.12 |
| NUCLEAR SCIENCE & TECHNOLOGY | 409 | 7.44 | 0.15 |
| NURSING | 492 | 7.19 | 0.04 |
| NUTRITION & DIETETICS | 635 | 7.20 | 0.02 |
| OBSTETRICS & GYNECOLOGY | 656 | 6.52 | 0.05 |
| OCEANOGRAPHY | 531 | 8.30 | 0.09 |
| ONCOLOGY | 1371 | 5.58 | 0.09 |
| OPERATIONS RESEARCH & MANAGEMENT SCIENCE | 618 | 8.60 | 0.06 |
| OPHTHALMOLOGY | 488 | 7.72 | 0.02 |
| OPTICS | 629 | 6.50 | 0.13 |
| ORNITHOLOGY | 204 | 9.77 | 0.02 |
| ORTHOPEDICS | 465 | 8.17 | 0.06 |
| OTORHINOLARYNGOLOGY | 320 | 8.45 | 0.01 |
| PALEONTOLOGY | 345 | 9.14 | 0.09 |
| PARASITOLOGY | 291 | 7.68 | 0.06 |
| PATHOLOGY | 833 | 7.15 | 0.14 |
| PEDIATRICS | 880 | 7.11 | 0.10 |
| PERIPHERAL VASCULAR DISEASE | 543 | 6.10 | 0.12 |
| PHARMACOLOGY & PHARMACY | 2037 | 6.60 | 0.09 |
| PHILOSOPHY | 54 | 7.47 | 0.21 |
| PHYSICS, APPLIED | 968 | 6.57 | 0.09 |
| PHYSICS, ATOMIC, MOLECULAR & CHEMICAL | 368 | 8.27 | 0.04 |
| PHYSICS, CONDENSED MATTER | 686 | 7.51 | 0.13 |
| PHYSICS, FLUIDS & PLASMAS | 302 | 7.42 | 0.21 |
| PHYSICS, MATHEMATICAL | 393 | 7.35 | 0.18 |
| PHYSICS, MULTIDISCIPLINARY | 707 | 7.99 | 0.10 |
| PHYSICS, NUCLEAR | 246 | 6.45 | 0.10 |
| PHYSICS, PARTICLES & FIELDS | 201 | 5.93 | 0.12 |

Table 1. (con't)

| CATEGORY | Journals | Mean Cited Half-Life | Growth per Annum |
|---|---|---|---|
| PHYSIOLOGY | 791 | 7.25 | 0.05 |
| PLANNING & DEVELOPMENT | 534 | 7.64 | 0.14 |
| PLANT SCIENCES | 1837 | 8.31 | 0.11 |
| POLITICAL SCIENCE | 996 | 7.95 | 0.15 |
| POLYMER SCIENCE | 612 | 7.45 | 0.04 |
| PRIMARY HEALTH CARE | 31 | 8.04 | 0.44 |
| PSYCHIATRY | 1653 | 7.12 | 0.06 |
| PSYCHOLOGY | 1016 | 8.47 | 0.10 |
| PSYCHOLOGY, APPLIED | 767 | 8.47 | 0.11 |
| PSYCHOLOGY, BIOLOGICAL | 188 | 8.66 | 0.02 |
| PSYCHOLOGY, CLINICAL | 1131 | 8.13 | 0.07 |
| PSYCHOLOGY, DEVELOPMENTAL | 710 | 8.68 | 0.04 |
| PSYCHOLOGY, EDUCATIONAL | 601 | 8.77 | 0.01 |
| PSYCHOLOGY, EXPERIMENTAL | 908 | 8.94 | 0.00 |
| PSYCHOLOGY, MATHEMATICAL | 160 | 10.06 | 0.05 |
| PSYCHOLOGY, MULTIDISCIPLINARY | 1130 | 8.51 | 0.06 |
| PSYCHOLOGY, PSYCHOANALYSIS | 167 | 9.54 | 0.13 |
| PSYCHOLOGY, SOCIAL | 637 | 9.36 | 0.07 |
| PUBLIC ADMINISTRATION | 367 | 7.49 | 0.10 |
| PUBLIC, ENVIRONMENTAL & OCCUPATIONAL HEALTH | 1549 | 7.20 | 0.10 |
| RADIOLOGY, NUCLEAR MEDICINE & MEDICAL IMAGING | 1084 | 6.70 | 0.08 |
| REHABILITATION | 741 | 7.70 | 0.04 |
| REMOTE SENSING | 119 | 7.56 | 0.11 |
| REPRODUCTIVE BIOLOGY | 271 | 6.36 | 0.16 |
| RESPIRATORY SYSTEM | 370 | 6.44 | 0.15 |
| RHEUMATOLOGY | 197 | 6.35 | 0.00 |
| ROBOTICS | 106 | 6.38 | -0.03 |
| SOCIAL ISSUES | 426 | 7.51 | 0.07 |
| SOCIAL SCIENCES, BIOMEDICAL | 272 | 7.38 | 0.07 |
| SOCIAL SCIENCES, INTERDISCIPLINARY | 831 | 8.74 | 0.00 |
| SOCIAL SCIENCES, MATHEMATICAL METHODS | 374 | 9.90 | 0.12 |
| SOCIAL WORK | 386 | 7.93 | 0.05 |
| SOCIOLOGY | 1372 | 9.14 | 0.10 |
| SOIL SCIENCE | 380 | 8.33 | 0.09 |
| SPECTROSCOPY | 486 | 6.94 | 0.11 |
| SPORT SCIENCES | 529 | 8.20 | 0.05 |
| STATISTICS & PROBABILITY | 805 | 9.43 | 0.09 |
| SUBSTANCE ABUSE | 322 | 6.56 | 0.11 |
| SURGERY | 1555 | 7.55 | 0.07 |
| TELECOMMUNICATIONS | 393 | 7.03 | 0.02 |
| THERMODYNAMICS | 491 | 7.57 | 0.01 |
| TOXICOLOGY | 792 | 6.88 | 0.05 |
| TRANSPLANTATION | 157 | 6.01 | 0.16 |
| TRANSPORTATION | 202 | 8.01 | -0.01 |
| TRANSPORTATION SCIENCE & TECHNOLOGY | 188 | 7.47 | -0.05 |

Table 1. (con't)

| CATEGORY | Journals | Mean Cited Half-Life | Growth per Annum |
|---|---|---|---|
| TROPICAL MEDICINE | 170 | 7.77 | 0.12 |
| URBAN STUDIES | 422 | 7.46 | 0.20 |
| UROLOGY & NEPHROLOGY | 476 | 6.02 | 0.10 |
| VETERINARY SCIENCES | 1266 | 8.18 | 0.05 |
| VIROLOGY | 306 | 6.17 | 0.18 |
| WATER RESOURCES | 423 | 7.73 | 0.11 |
| WOMEN'S STUDIES | 320 | 8.35 | 0.08 |
| ZOOLOGY | 1382 | 9.21 | 0.05 |